# Dynamic Modeling of Magneto-Active Grounding Electrodes under Transient Conditions

José M. Campos-Salazar
*Electronic Engineering Department)*
*Universitat Politècnica de Catalunya*
Barcelona, Spain
jose.manuel.campos@upc.edu

***Abstract*—Grounding systems operating under transient electromagnetic conditions exhibit highly nonlinear behavior strongly influenced by electromagnetic propagation, soil conductivity variations, thermal diffusion, moisture transport, and ionization phenomena. Conventional grounding analyses generally rely on static resistance formulations that neglect the coupled interactions governing the transient response of advanced grounding technologies. In particular, rigorous dynamic models describing magneto-active grounding electrode (MAGE) systems remain practically nonexistent within the current scientific literature despite the increasing commercial interest in grounding technologies incorporating electromagnetic coupling structures and conductive enhancement compounds. A comprehensive nonlinear dynamic model for a MAGE system is therefore developed considering electromagnetic coupling, dynamic grounding impedance behavior, conductivity adaptation, electrothermal effects, moisture-dependent soil dynamics, and nonlinear ionization mechanisms. The proposed formulation combines transient electromagnetic equations, thermal energy balances, conductive adaptation models, magnetic flux dynamics, and soil ionization representations into a unified MATLAB/Simulink framework. Numerical simulations were performed under combined step and impulsive transient excitation conditions to evaluate the interaction among the coupled subsystems. The obtained results demonstrate that the equivalent grounding impedance evolves dynamically according to environmental and electromagnetic operating conditions, confirming that grounding systems cannot be accurately represented through constant-parameter formulations under transient conditions. Furthermore, the simulations reveal stable nonlinear coupling among electromagnetic propagation, conductivity enhancement, thermal diffusion, moisture redistribution, and magneto-active interaction phenomena. The proposed framework constitutes a proof-of-concept analytical basis for future transient analysis and optimization of advanced grounding technologies incorporating magnetically coupled grounding structures.**

## I. INTRODUCTION

Grounding systems constitute one of the most critical subsystems in modern electrical power installations due to their direct influence on operational safety, electromagnetic compatibility, equipment protection, transient overvoltage mitigation, lightning current dissipation, insulation coordination, and fault-current management [1]–[5]. The principal objective of a grounding system is to provide a low-impedance conductive path between electrical installations and the surrounding earth medium, thereby enabling safe dissipation of fault currents, leakage currents, switching disturbances, and atmospheric discharge phenomena into the soil structure [5]–[7]. Consequently, the electrical and transient behavior of grounding systems has become increasingly important in modern power systems characterized by elevated short-circuit levels, high-frequency switching devices, power-electronic converters, distributed energy resources, renewable generation systems, and complex electromagnetic operating environments [2]–[4], [8]–[11].

Traditionally, grounding systems have been modeled using simplified static formulations based on constant soil resistivity and lumped grounding resistance assumptions [5]–[7]. Classical grounding analyses generally represent the grounding electrode as a purely resistive element whose electrical behavior is assumed invariant with respect to current magnitude, transient excitation, frequency content, thermal effects, moisture dynamics, and environmental operating conditions. Although such simplified formulations may provide acceptable approximations under steady-state low-frequency operating conditions, they become increasingly inaccurate under transient electromagnetic phenomena such as lightning strikes, impulsive fault currents, switching surges, and high-frequency disturbances [2], [3], [10]–[22].

Recent investigations have demonstrated that grounding systems exhibit highly nonlinear and frequency-dependent behavior under transient operating conditions [8]–[12], [15], [15], [19], [20]. In particular, several studies have shown that transient grounding performance is strongly influenced by electromagnetic wave propagation, capacitive coupling, inductive interactions, soil ionization, moisture-dependent conductivity variations, electrothermal effects, and environmental dynamics [2], [8], [9], [12], [17], [19], [21]–[23]. Consequently, the equivalent grounding impedance cannot be accurately interpreted as a constant parameter, but rather as a dynamic quantity evolving according to coupled electromagnetic and physicochemical interactions occurring within the soil–electrode interface region [8], [12], [15], [19], [24], [25], [25]–[29].

The increasing complexity of modern electrical systems has motivated the development of advanced grounding technologies aimed at improving transient dissipation capability and reducing equivalent grounding impedance. Among these technologies, conductive backfill compounds, soil-conditioning materials, frequency-dependent grounding topologies, and transient-optimized electrode structures have received substantial research attention in recent years [1], [30]–[33]. Conductive enhancement materials based on bentonite, graphite, ionic compounds, and hygroscopic agents have been widely investigated to improve long-term grounding stability and reduce seasonal grounding resistance variations [1], [31]–[33]. Similarly, frequency-dependent grounding models and distributed electromagnetic formulations have been proposed to improve the representation of transient current propagation and lightning-induced phenomena [2], [10]–[12], [15], [19]–[22].

Additionally, recent studies have investigated the influence of thermal phenomena, electrothermal coupling, and soil ionization effects on grounding performance under high-current conditions [8], [9], [28], [29], [34]–[36]. These investigations demonstrated that substantial transient current injection may produce localized Joule heating, conductivity enhancement, and nonlinear soil ionization effects capable of significantly modifying the equivalent grounding impedance and current dissipation characteristics [27]–[29], [34]–[36]. Furthermore, moisture transport dynamics and seasonal soil variations have been identified as critical factors influencing grounding-system reliability and long-term operational stability [23], [32], [37].

Despite the significant progress achieved in the fields of transient grounding analysis and nonlinear grounding-system modeling, the available literature remains primarily focused on conventional passive grounding structures [2], [8]–[12], [15], [19]–[22], [24]–[29], [38]. In particular, most existing grounding models are based on metallic rods, grounding grids, conductive meshes, counterpoises, or conductive backfill systems operating under passive conductive dissipation mechanisms. Although these studies have considerably improved the understanding of grounding-system dynamics, the integration of active or magnetically coupled subsystems into grounding electrodes has received extremely limited attention in the scientific literature.

In recent years, commercial grounding technologies incorporating magneto-active assemblies, electromagnetic modules, and hybrid conductive structures have emerged as alternative grounding solutions intended to enhance transient current dispersion and reduce equivalent grounding impedance under impulsive operating conditions. One representative example corresponds to the magneto-active grounding electrode systems commercialized and implemented by [39], which integrate conductive electrodes, conductive backfill compounds, geometric field-enhancement structures, and magnetically coupled subsystems into a unified grounding topology [39]. However, despite the growing commercial interest in such technologies, it is currently not possible to find rigorous dynamic mathematical models describing the coupled electromagnetic, thermal, conductive, and moisture-dependent behavior of magneto-active grounding electrode (MAGE) systems within the scientific literature.

More specifically, no detailed nonlinear dynamic models describing the transient interaction among electromagnetic coupling, soil conductivity dynamics, thermal diffusion, moisture transport, grounding potential rise, and ionization effects associated with magneto-active grounding electrodes were identified during the literature review performed for this work. Similarly, no state-space formulations, transient electromagnetic models, or coupled dynamic simulation studies addressing the transient operation of MAGE systems under fault or impulsive excitation conditions were found. Therefore, the absence of analytical and numerical modeling frameworks for these grounding technologies constitutes an important scientific gap within the current grounding-system literature.

The development of such models becomes particularly relevant because magneto-active grounding structures cannot be interpreted as purely passive conductive systems. Instead, these topologies exhibit coupled electromagnetic interactions introduced by the magneto-active assembly, conductive adaptation associated with the surrounding backfill material, thermal diffusion phenomena caused by transient current dissipation, and moisture-dependent conductivity variations occurring within the soil medium. Consequently, the resulting grounding behavior becomes inherently nonlinear, distributed, and in nature.

Under transient excitation conditions, the simultaneous interaction among conductive dissipation, magnetic coupling, capacitive polarization, thermal diffusion, and soil ionization may substantially modify the equivalent grounding impedance and transient current propagation characteristics. Therefore, the use of classical static grounding formulations becomes insufficient to accurately represent the physical behavior of these systems under realistic operating conditions.

Motivated by the aforementioned limitations, this work proposes a comprehensive nonlinear dynamic model for a magneto-active grounding electrode system considering electromagnetic coupling, soil conductivity dynamics, thermal effects, moisture transport, and ionization phenomena. The proposed formulation combines electromagnetic transient equations, dynamic grounding impedance models, thermal energy balances, moisture transport dynamics, conductivity adaptation mechanisms, and magneto-active coupling into a unified framework implemented in the MATLAB/Simulink environment.

Unlike conventional grounding analyses based on static resistance assumptions, the proposed approach models the grounding electrode as a dynamic distributed subsystem whose equivalent impedance evolves according to environmental conditions, transient electromagnetic interactions, and conductive adaptation mechanisms. The proposed model additionally incorporates a magneto-active correction subsystem intended to represent the electromagnetic influence introduced by the magnetic assembly integrated into the grounding structure.

It is important to emphasize that the present study corresponds exclusively to a theoretical proof-of-concept investigation focused on the development and validation of the proposed mathematical framework through numerical simulations. Consequently, no experimental validation, field measurements, laboratory testing, or prototype implementation has been carried out within the scope of this work. The principal objective of the manuscript is therefore not to experimentally validate the grounding technology itself, but rather to establish an initial analytical and computational framework capable of representing the transient behavior potentially associated with magneto-active grounding electrodes.

The proposed proof-of-concept approach enables the preliminary evaluation of the dynamic interactions among electromagnetic propagation, conductivity adaptation, thermal diffusion, moisture transport, and nonlinear grounding impedance evolution under transient operating conditions. Furthermore, the developed framework may constitute a foundational basis for future experimental investigations, transient grounding optimization studies, advanced electromagnetic simulations, and subsequent validation campaigns associated with next-generation grounding technologies.

The remainder of the manuscript is organized as follows. Section 2 describes the topology and principal constructive characteristics of the proposed magneto-active grounding electrode system. Section 3 develops the complete nonlinear dynamic model, including the electromagnetic transient equations, conductivity dynamics, thermal and moisture models, magneto-active subsystem, and soil ionization formulation. Section 4 presents the MATLAB/Simulink implementation and analyzes the transient simulation results obtained under combined step and impulsive excitation conditions. Finally, the principal conclusions and future research directions are summarized in Section 5.

## II. System Topology

The MAGE system corresponds to an advanced grounding topology specifically developed to enhance the dissipation capability of electrical fault currents, leakage currents, switching disturbances, and high-energy transient phenomena into the surrounding soil medium. Unlike conventional grounding electrodes, which mainly rely on passive conductive dissipation through metallic rods or meshes, the magneto-active electrode integrates conductive, electromagnetic, geometric, and physicochemical mechanisms into a single hybrid grounding structure. This integrated topology enables improved transient current dispersion, reduction of equivalent grounding impedance, enhancement of soil ion mobility, and mitigation of overvoltage propagation under both steady-state and transient operating conditions [1]–[5].

Fig. 1 illustrates the proposed topology of the magneto-active grounding electrode system, including its principal constructive and functional components.

The topology generally consists of a vertically installed conductive electrode manufactured from high-purity electrolytic copper or copper-based alloys, mechanically coupled with a magneto-active module and embedded within a conductive backfill medium specifically formulated to reduce soil resistivity. The grounding structure is designed to operate as a coupled conductive–electromagnetic interface between the electrical installation and the surrounding earth medium. Consequently, the electrode cannot be interpreted exclusively as a purely resistive grounding element, but rather as a distributed subsystem exhibiting electromagnetic, thermal, electrochemical, and moisture-dependent interactions [2], [8]–[11], [40]–[42].

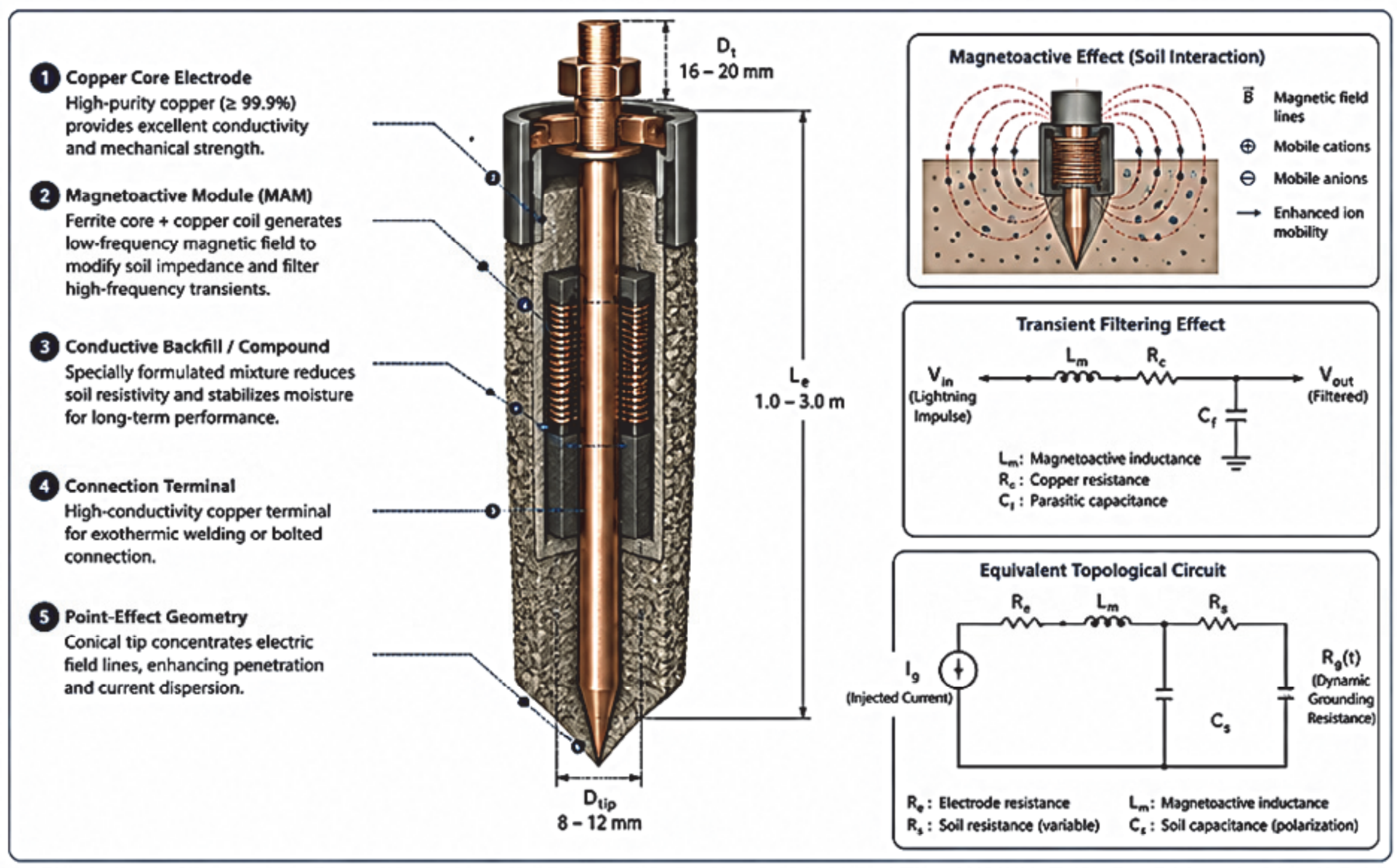

Fig. 1. System topology of the magnetoactive grounding electrode developed by [39]. The system integrates copper core, magnetoactive module, conductive backfill, and point-effect geometry to optimize grounding performance under steady-state and transient conditions.

From a topological perspective, the system incorporates five principal functional subsystems:

1. Primary conductive electrode.
2. Magneto-active electromagnetic module.
3. Conductive soil-conditioning compound.
4. Mechanical and electrical connection terminal.
5. Geometric field-concentration structure.

The interaction among these subsystems generates a synergistic grounding behavior that differs substantially from that of conventional grounding rods. In particular, the magneto-active assembly introduces electromagnetic coupling effects capable of modifying the local current distribution and transient propagation characteristics within the soil–electrode interface region [2], [10], [11], [30], [31].

Furthermore, the conductive backfill compound surrounding the electrode contributes to stabilizing soil resistivity through moisture retention and ionic enhancement mechanisms. This feature becomes particularly important in soils exhibiting large seasonal variations in humidity, temperature, salinity, and compaction conditions [12], [23], [32], [37]. Consequently, the grounding performance of the system becomes strongly dependent on environmental variables and soil dynamic properties.

As can be seen in Fig. 1, the primary conductive electrode constitutes the main electrical current injection path between the protected electrical system and the surrounding earth medium. In most practical implementations, the electrode is manufactured using electrolytic copper with purity levels exceeding 99.9%, owing to its high electrical conductivity, corrosion resistance, thermal stability, and long-term operational reliability [43].

The electrode geometry is commonly designed using elongated cylindrical or conical structures (see Figure 1). In particular, conical tip geometries are frequently employed to intensify local electric field concentration effects near the terminal region of the electrode. This so-called point-effect phenomenon promotes improved electric field penetration into the soil medium, facilitating enhanced transient current dispersion and reduced effective grounding impedance during high-energy electrical disturbances [6], [7].

The physical dimensions of the electrode (see Figure 1), including length, diameter, insertion depth, and geometric aspect ratio, strongly influence the resulting grounding resistance and transient impedance characteristics. Larger electrode lengths generally increase the effective contact area with the soil, thereby improving current dissipation capability. Similarly, increased insertion depth allows access to deeper soil layers with more stable moisture content and lower resistivity [13], [42].

Additionally, the conductive electrode exhibits distributed electrical behavior due to parasitic resistance, capacitance, and inductive coupling effects associated with transient current propagation. Therefore, under fast transient conditions such as lightning impulses or switching surges, the electrode cannot be modeled as a lumped resistance alone, since frequency-dependent electromagnetic effects become significant [12], [14]–[16].

One of the principal distinguishing characteristics of the proposed topology is the incorporation of a magneto-active electromagnetic module integrated around the conductive electrode structure. This subsystem generally consists of a magnetic core assembly combined with conductive windings or inductive structures capable of generating localized magnetic fields during transient current propagation.

The magneto-active module introduces electromagnetic coupling phenomena into the soil–electrode interface region. From a physical perspective, the time-varying magnetic field generated around the electrode may influence ionic mobility, local current dispersion patterns, and transient electromagnetic propagation characteristics within the surrounding conductive medium [18], [31], [44].

Additionally, the electromagnetic module behaves as a distributed filtering subsystem capable of attenuating high-frequency transient components associated with lightning discharges, switching overvoltages, and impulsive fault currents. Consequently, the grounding system exhibits not only conductive dissipation behavior but also partial transient electromagnetic filtering characteristics [19]–[22].

The magnetic assembly may incorporate ferrite materials, laminated magnetic cores, or hybrid inductive structures designed to operate under broadband frequency conditions. The resulting electromagnetic behavior becomes highly dependent on parameters such as magnetic permeability, core saturation characteristics, winding topology, transient current amplitude, and electromagnetic coupling between the electrode and surrounding soil medium [45]–[47].

Moreover, the presence of the magneto-active subsystem introduces nonlinear electromagnetic phenomena due to magnetic saturation, hysteresis effects, and frequency-dependent magnetic permeability variations. These phenomena become especially relevant under high-energy transient conditions where large current magnitudes produce significant electromagnetic field intensification [48], [49].

The conductive backfill or soil-conditioning compound constitutes another critical subsystem within the grounding topology. This material is typically composed of bentonite, graphite, conductive minerals, hygroscopic compounds, and ionic agents specifically formulated to reduce the equivalent resistivity of the surrounding soil medium [31], [50], [51].

The principal function of the conductive compound is to enhance the electrical contact between the electrode surface and the earth medium. This enhancement is achieved through three principal mechanisms:

1. Reduction of local soil resistivity.
2. Stabilization of soil moisture content.
3. Increase of ionic conductivity around the electrode interface.

The conductive backfill also contributes to improving long-term grounding stability by minimizing seasonal variations in grounding resistance caused by environmental changes such as drought periods, rainfall cycles, freezing conditions, and thermal fluctuations [37], [52].

Furthermore, the conductive compound exhibits coupled thermal–electrical behavior. During transient current dissipation events, Joule heating effects may alter the local moisture content and conductivity distribution surrounding the electrode. Consequently, the grounding system becomes inherently dynamic due to the interaction between thermal diffusion, moisture transport, and conductivity variations [36], [53], [54].

In highly energetic transient events, soil ionization phenomena may also appear around the electrode region. Such ionization processes can temporarily reduce the effective grounding impedance and significantly modify current propagation characteristics within the earth medium [9], [42], [55].

The grounding electrode incorporates a high-conductivity terminal interface designed to ensure reliable electrical continuity between the grounding system and the protected electrical installation. This connection may be implemented using exothermic welding techniques, bolted copper terminals, compression connectors, or hybrid conductive interfaces.

The connection subsystem must satisfy stringent electrical, thermal, and mechanical requirements due to the large transient currents potentially flowing through the grounding structure during fault events. Poor connection quality may generate localized overheating, contact resistance growth, accelerated corrosion, and degradation of grounding performance over time [56]–[59].

Mechanical protection structures may additionally be incorporated to improve installation robustness and protect the grounding system against environmental degradation, soil displacement, mechanical vibrations, and external impacts [60].

The overall grounding topology behaves as a coupled conductive–electromagnetic–thermal–soil interaction system. Consequently, the grounding electrode exhibits dynamic behavior strongly influenced by soil properties, transient current waveforms, environmental conditions, and electromagnetic coupling mechanisms.

Under steady-state operating conditions, the system primarily dissipates low-frequency leakage and fault currents through conductive mechanisms governed by soil resistivity and electrode geometry. However, under transient conditions such as lightning impulses or switching overvoltages, the grounding system exhibits substantially more complex behavior due to the simultaneous presence of:

frequency-dependent impedance effects,

transient electromagnetic propagation,

distributed capacitive coupling,

inductive interactions,

soil ionization,

thermal diffusion,

and moisture-dependent conductivity variations [15], [19], [61]–[64].

Therefore, the proposed grounding topology must be interpreted as a nonlinear distributed system rather than a purely resistive grounding element. This observation motivates the development of advanced dynamic models capable of representing the coupled electromagnetic, thermal, hydraulic, and electrochemical phenomena governing the transient behavior of magneto-active grounding electrodes.

## III. System Modeling

### A. General Modeling Framework

The proposed MAGE system corresponds to a highly coupled nonlinear distributed structure whose transient behavior is governed by simultaneous electromagnetic, conductive, thermal, electrochemical, and moisture-dependent phenomena occurring within the soil–electrode interface region. Consequently, the grounding system cannot be accurately represented using classical static grounding resistance formulations exclusively, particularly under transient fault currents, lightning discharges, impulsive overvoltages, or fast electromagnetic disturbances [12], [14]–[16], [18], [31], [44].

The complete system topology exhibits dynamic interactions among:

the conductive copper electrode,

the magneto-active electromagnetic module,

the surrounding conductive backfill medium,

the soil ionization region,

the thermal diffusion mechanism,

and the moisture-dependent conductivity distribution.

Therefore, the grounding structure must be modeled as a nonlinear distributed dynamic system composed of coupled electrical, electromagnetic, thermal, and hydraulic subsystems [19]–[22], [45]–[47].

The proposed modeling framework considers the following principal state variables, i.e., $i_g(t)$ which is the grounding current injected into the soil. Also, the equivalent electrode potential rise is denoted as $v_g(t)$. Furthermore, $T_s(t)$, $\theta_w(t)$, and $\sigma_s(t)$ are the equivalent soil temperature surrounding the electrode, the equivalent volumetric soil moisture content, and the equivalent soil conductivity, respectively. Finally, the magnetic flux associated with the magneto-active module and the equivalent dynamic grounding resistance are denoted as $\Phi_m(t)$ and $R_g(t)$, respectively.

The resulting model combines:

1. Electromagnetic transient equations,
2. Dynamic grounding impedance behavior,

3. Conductive soil interaction,
4. Thermal diffusion,
5. Moisture transport,
6. Soil ionization effects.

## B. Electromagnetic Equivalent Model

The transient electrical behavior of the grounding electrode can be represented through an equivalent distributed electromagnetic topology composed of resistive, inductive, and capacitive components associated with the electrode, surrounding soil, and magneto-active module [31], [48]–[51], [51], [65]. This equivalent topology is shown in Fig. 2.

The transient current propagation through the grounding system is governed by:

$$v_f(t) = v_L(t) + v_R(t) + v_g(t) \tag{1}$$

where:

$$v_L(t) = L_e \cdot \frac{di_g(t)}{dt} \tag{2}$$

corresponds to the electrode inductive voltage,

$$v_R(t) = R_e \cdot i_g(t) \tag{3}$$

represents conductive losses within the electrode, and $v_g(t)$ is the equivalent grounding potential rise. Thus:

$$v_L(t) = L_e \cdot \frac{di_g(t)}{dt} + R_e \cdot i_g(t) + v_g(t) \tag{4}$$

where, the injected fault or transient voltage in V; the equivalent electrode inductance in H; the equivalent electrode resistance in Ω; and the grounding current in A are denoted as $v_f(t)$; $L_e$; $R_e$; and $i_g(t)$, respectively.

The grounding potential dynamics are modeled as:

$$C_s \cdot \frac{dv_g(t)}{dt} = i_g(t) + i_d(t) \tag{5}$$

where $C_s$ is the equivalent soil capacitance in F and $i_d(t)$ is the dissipated current through the surrounding soil in A.

The dissipated current is expressed as:

$$i_d(t) = v_g(t)/R_g(t) \tag{6}$$

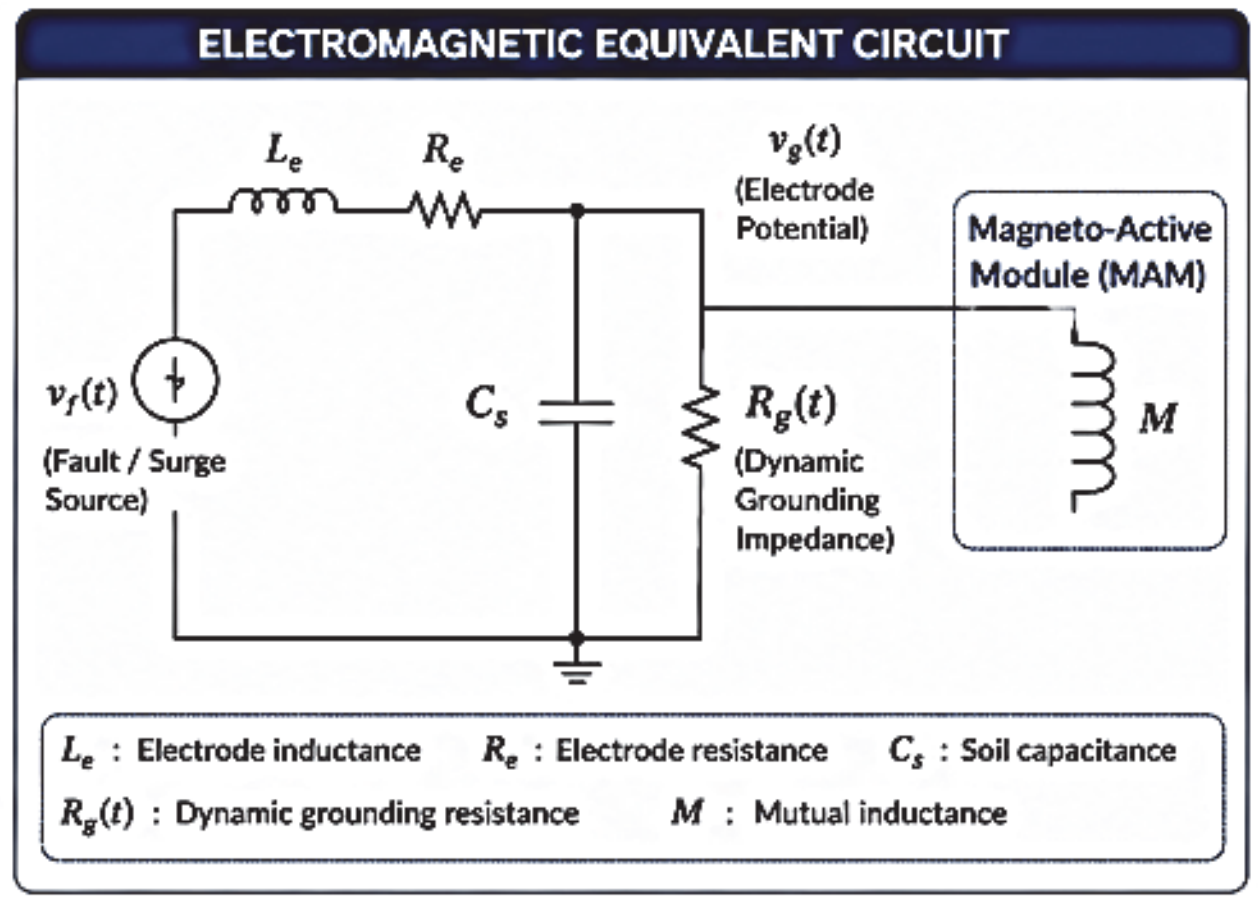

Fig. 2. Electromagnetic equivalent circuit.

Therefore

$$C_s \cdot \frac{dv_g(t)}{dt} = i_g(t) + \frac{v_g(t)}{R_g(t)} \tag{7}$$

This equation demonstrates that the grounding system behaves dynamically due to capacitive polarization and time-varying grounding impedance effects [31], [37], [37], [52], [65].

## C. Dynamic Grounding Resistance Model

Unlike classical grounding systems, the equivalent grounding resistance of the proposed topology is not constant. Instead, it depends on multiple coupled environmental and electromagnetic variables including:

soil conductivity,

temperature,

moisture content,

transient current density,

soil ionization,

and conductive backfill conditions.

Consequently, $R_g(t) = f(\sigma_s(t), T_s(t), \theta_w(t), J(t), \Phi_m(t))$. According to the Figure 1, the equivalent grounding resistance may be approximated as:

$$R_g(t) \approx \frac{1}{2 \cdot \pi \cdot L_e} \cdot \rho_s(t) \cdot \ln\left(\frac{4 \cdot L_e}{d_e}\right) \cdot K_m(t) \cdot K_i(t) \tag{8}$$

From here, $\rho_s(t)$, $L_e$, $d_e$, $K_m(t)$, and $K_i(t)$ are the dynamic soil resistivity in Ω·m, the electrode length in m, the electrode diameter in m, the magneto-active correction factor, and the ionization correction factor, respectively.

The dynamic soil resistivity is related to conductivity by:

$$\rho_s(t) = 1/\sigma_s(t) \tag{9}$$

## D. Soil Conductivity Dynamics

The conductivity of the surrounding soil medium varies dynamically due to thermal effects, moisture transport, ionic mobility, and electromagnetic interaction [9], [36], [53], [54].

The conductivity dynamics are represented as indicated in (10) where $k_\theta$, $k_T$, $k_m$, and $k_\sigma$ are the moisture conductivity coefficient, the thermal conductivity coefficient, the magnetic

$$\frac{d\sigma_s(t)}{dt} = k_\theta \cdot \theta_w(t) + k_T \cdot T_s(t) + k_m \cdot \Phi_m(t) - k_\sigma \cdot \sigma_s(t) \tag{10}$$

enhancement coefficient, and the conductivity dissipation coefficient, respectively.

This equation represents the dynamic conductive adaptation of the soil–electrode interface.

## E. Moisture Transport Dynamics

The conductive backfill material exhibits moisture-retention properties capable of stabilizing the equivalent grounding impedance. However, the local moisture distribution changes dynamically due to evaporation, thermal diffusion, and environmental conditions [42], [55].

The equivalent moisture dynamics are modeled as:

$$\frac{d\theta_W(t)}{dt} = q_{in}(t) - q_{ev}(t) - q_{dr}(t) \quad (11)$$

From here, $q_{in}(t)$, $q_{ev}(t)$, and $q_{dr}(t)$ are the moisture absorption rate, the evaporation rate, and the drainage/diffusion losses, respectively.

The evaporation dynamics may be approximated by:

$$q_{ev}(t) = k_{ev} \cdot (T_s(t) - T_{amb}) \quad (12)$$

where $k_{ev}$ is the evaporation coefficient and $T_{amb}$ is the ambient temperature in K.

Thus:

$$\frac{d\theta_W(t)}{dt} = q_{in}(t) - k_{ev} \cdot (T_s(t) - T_{amb}) - q_{dr}(t) \quad (13)$$

### F. Thermal Dynamic Model

Under transient high-current conditions, substantial Joule heating appears within the electrode and surrounding soil medium [56]–[61].

The thermal energy balance is expressed as:

$$C_{th} \cdot \frac{dT_S(t)}{dt} = P_J(t) - P_{loss}(t) \quad (14)$$

From here, $C_{th}$, $P_J(t)$, and $P_{loss}(t)$ are the equivalent thermal capacitance in J/K, the Joule heating power in W, and the thermal dissipation losses in W, respectively.

The Joule heating term is:

$$P_J(t) = R_g(t) \cdot i_g^2(t) \quad (15)$$

The thermal dissipation is:

$$P_{loss}(t) = h_s \cdot A_s \cdot (T_s(t) - T_{amb}) \quad (16)$$

From here, $h_s$ is the thermal convection coefficient and $A_s$ is the effective thermal exchange area, respectively.

Therefore:

$$C_{th} \cdot \frac{dT_S(t)}{dt} = R_g(t) \cdot i_g{}^2(t) - h_s \cdot A_s \cdot (T_s(t) - T_{amb}) \quad (17)$$

This equation describes the transient thermal evolution of the soil region surrounding the electrode.

### G. Magneto-Active Electromagnetic Model

According to the Fig. 2, the magneto-active module introduces localized electromagnetic coupling effects around the electrode region. The magnetic subsystem may be represented using an equivalent magnetic circuit model [15], [19], [62]–[64].

The magnetic flux dynamics are:

$$v_m(t) = R_m(t) \cdot \Phi_m(t) + N \cdot \frac{d\Phi_m(t)}{dt} \quad (18)$$

where the induced magnetic voltage, the magnetic reluctance, the number of turns, and the magnetic flux in Wb are denoted as $v_m(t)$, $R_m$, $N$, and $\Phi_m(t)$, respectively.

The electromagnetic coupling between the grounding current and magnetic subsystem is represented by:

$$v_m(t) = M \cdot \frac{di_g(t)}{dt} \quad (19)$$

In this case, M is the mutual inductance coefficient.

Therefore:

$$M \cdot \frac{di_g(t)}{dt} = R_m(t) \cdot \Phi_m(t) + N \cdot \frac{d\Phi_m(t)}{dt} \quad (20)$$

This equation characterizes the dynamic electromagnetic interaction between transient grounding currents and the magneto-active assembly.

### H. Soil Ionization Model

Under high-current impulsive conditions, the electric field surrounding the electrode may exceed the critical soil ionization threshold, causing nonlinear conductivity enhancement [9], [11], [14], [66], [67].

## IV. Simulation Results

The proposed dynamic model of the MAGE system was implemented and simulated in the MATLAB/Simulink environment using a modular structure composed of coupled electrical, electromagnetic, thermal, conductivity, moisture, and soil-ionization subsystems. The complete simulation framework incorporates the nonlinear dynamic equations developed in Section 3, including the electromagnetic transient model, dynamic grounding resistance formulation, thermal diffusion equations, moisture transport dynamics, magneto-active coupling, and nonlinear ionization correction mechanisms. The implemented Simulink structure is shown in the developed simulation architecture presented in the manuscript, where each subsystem was independently modeled and subsequently coupled through nonlinear interaction variables.

The transient excitation was introduced through a combined fault-voltage waveform composed of a step component and a double-exponential impulsive disturbance, enabling the evaluation of the grounding system under representative transient operating conditions associated with switching disturbances and moderate impulsive events. The numerical integration process was performed using variable-step solvers with sufficiently small integration steps to preserve numerical stability and accurately reproduce the fast transient dynamics associated with electromagnetic propagation within the soil–electrode interface region.

The complete simulation parameters employed throughout the numerical analysis are summarized in Table 1.

Fig. 3.1 presents the transient evolution of the grounding current and grounding potential rise generated during the disturbance injection process. The obtained results demonstrate that the proposed grounding system exhibits a stable transient response characterized by smooth current growth and progressive stabilization toward steady-state operating conditions. The grounding current initially experiences a rapid transient increase associated with the electromagnetic charging dynamics of the equivalent grounding subsystem. Subsequently, the current gradually converges toward a quasi-steady operating regime due to the combined effects of conductive dissipation and dynamic grounding impedance adaptation.

TABLE I. PRINCIPAL SIMULATION PARAMETERS USED IN THE MATLAB/SIMULINK IMPLEMENTATION

| Parameter | Description | Value |
|---|---|---|
| $L_e$ | Electrode length | 2 [m] |
| $d_e$ | Electrode diameter | 0.018 [m] |
| $R_e$ | Equivalent electrode resistance | 1.5 x $10^{-3}$ [Ω] |
| $L_{eq}$ | Equivalent electrode inductance | 5 x $10^{-6}$ [H] |
| $C_s$ | Equivalent soil capacitance | 5 x $10^{-6}$ [F] |
| $C_{th}$ | Equivalent thermal capacitance | 8 x $10^{4}$ [J/K] |
| $h_s$ | Thermal transfer coefficient | 5 [W/(m²·K)] |
| $\sigma_{s0}$ | Initial soil conductivity | 0.02 [S/m] |
| $\theta_{w0}$ | Initial volumetric moisture content | 0.2 |
| $N$ | Equivalent magnetic turns | 300 |
| $M$ | Mutual inductance coefficient | 2 x $10^{-5}$ [H] |
| $R_m$ | Magnetic reluctance coefficient | 20 |
| $E_{crit}$ | Critical ionization electric field | 3 x $10^{5}$ [V/m] |
| $V_{fault}$ | Step fault voltage amplitude | 1000 [V] |
| $V_{0\ imp}$ | Impulsive voltage amplitude | 5000 [V] |
| $\alpha_{imp}$ | Impulsive waveform coefficient | 2 x $10^{4}$ [1/s] |
| $\beta_{imp}$ | Impulsive waveform coefficient | 8 x $10^{5}$ [1/s] |

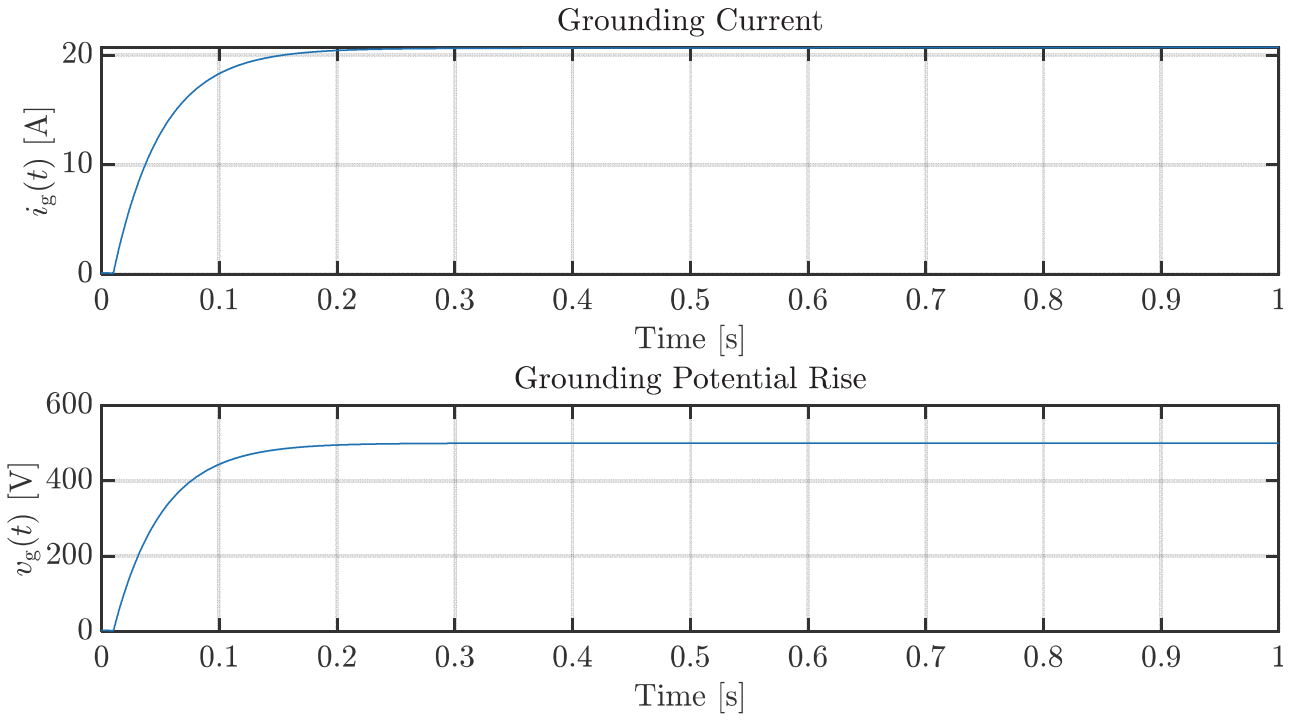

Fig. 3.1. Transient evolution of the grounding current $i_g(t)$ and grounding potential rise $v_g(t)$ under combined step and impulsive fault excitation. The results demonstrate the dynamic electromagnetic behavior of the grounding structure, including transient current propagation, capacitive polarization effects, and stabilization of the equivalent grounding potential during the disturbance dissipation process.

Simultaneously, the grounding potential rise exhibits a monotonic transient increase followed by asymptotic stabilization. This behavior confirms the presence of capacitive polarization dynamics within the soil–electrode interface region. The obtained transient response indicates that the grounding structure behaves as a distributed dynamic subsystem rather than a purely resistive grounding element. Furthermore, the absence of oscillatory instability demonstrates the numerical robustness and physical consistency of the proposed formulation.

The results shown in Fig. 3.2 correspond to the transient evolution of the equivalent grounding resistance and soil conductivity. The dynamic grounding resistance exhibits a gradual reduction throughout the transient process, confirming the nonlinear adaptive behavior of the grounding system. This impedance reduction is primarily associated with the combined influence of conductivity enhancement, electromagnetic coupling effects, and conductive adaptation occurring within the surrounding soil medium. Unlike classical grounding models assuming constant grounding resistance, the proposed formulation demonstrates that the grounding impedance evolves dynamically as a consequence of coupled environmental and electromagnetic interactions.

The conductivity results further validate this behavior. Specifically, the soil conductivity progressively increases during the transient process due to thermal interaction, electromagnetic coupling, and moisture-dependent conductive enhancement. The obtained conductivity dynamics remain smooth and numerically stable throughout the simulation interval, confirming the physical coherence of the proposed nonlinear conductivity formulation.

Fig. 3.3 illustrates the thermal and moisture dynamics associated with the grounding electrode system. The soil temperature exhibits a slow transient increase caused by Joule heating generated during current dissipation through the grounding structure. The thermal evolution remains physically realistic, presenting moderate temperature variations without excessive thermal amplification. This behavior indicates that the equivalent thermal capacitance and thermal dissipation mechanisms adequately regulate the thermal energy accumulation process.

The volumetric soil moisture content exhibits a slight gradual reduction throughout the transient interval due to evaporation effects coupled with the temperature increase.

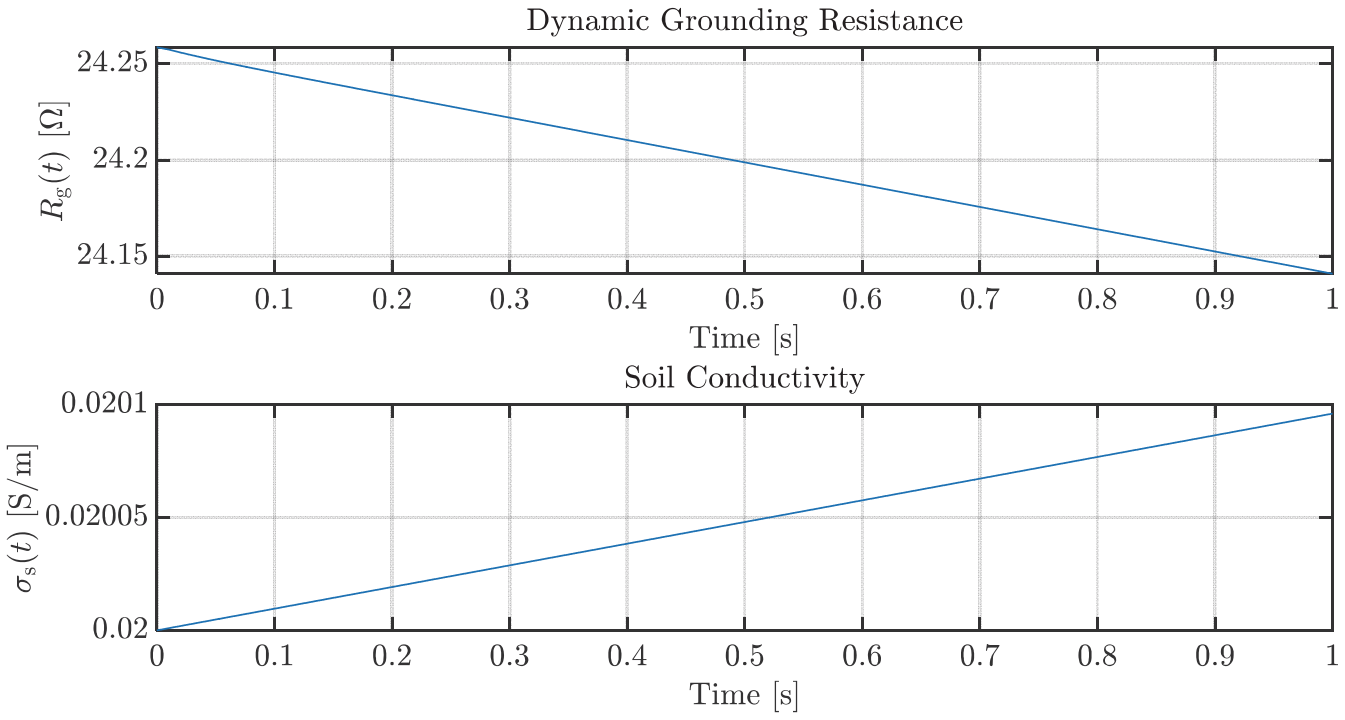

Fig. 3.2. Dynamic behavior of the equivalent grounding resistance $R_g(t)$ and soil conductivity $\sigma_s(t)$ during transient operation of the proposed magneto-active grounding electrode system. The results illustrate the nonlinear coupling between conductive adaptation and transient electromagnetic interactions, demonstrating the gradual reduction of grounding impedance caused by conductivity enhancement within the soil–electrode interface region.

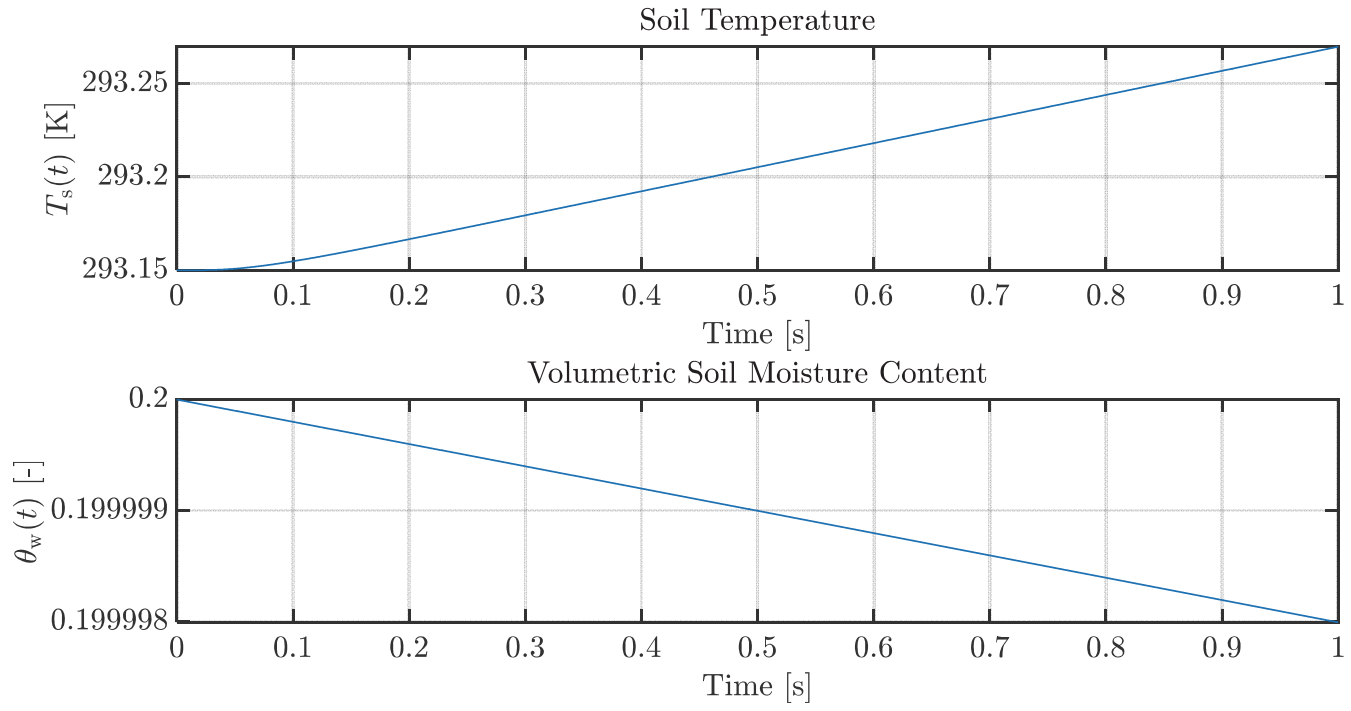

Fig. 3.3. Transient thermal and moisture responses of the proposed magneto-active grounding electrode system, showing the evolution of the soil temperature $T_s(t)$ and volumetric soil moisture content $\theta_w(t)$ under electromagnetic disturbance conditions. The results demonstrate the coupled electrothermal and moisture-transport dynamics associated with Joule heating, thermal diffusion, evaporation effects, and conductive stabilization within the soil–electrode interface region..

However, the moisture variation remains relatively small, demonstrating that the conductive backfill and surrounding soil medium maintain stable moisture-retention characteristics under the considered operating conditions. This result further confirms the long-term stability potential of the proposed grounding topology under moderate transient excitation.

The electromagnetic behavior of the magneto-active subsystem is presented in Fig. 3.4. The magnetic flux generated within the magneto-active assembly exhibits a stable transient evolution characterized by rapid initial growth followed by gradual stabilization. The obtained magnetic flux magnitude remains within physically realistic ranges, indicating that the magnetic subsystem operates below saturation conditions under the considered transient excitation.

Additionally, the magneto-active correction factor exhibits a slight reduction during the transient process. This behavior confirms that the electromagnetic coupling mechanism contributes to modifying the equivalent transient grounding response. Although the obtained correction remains moderate, the results demonstrate the capability of the magneto-active module to dynamically influence the grounding behavior through electromagnetic interaction effects. Importantly, the obtained response remains smooth and numerically stable, confirming the robustness of the proposed magnetic coupling formulation.

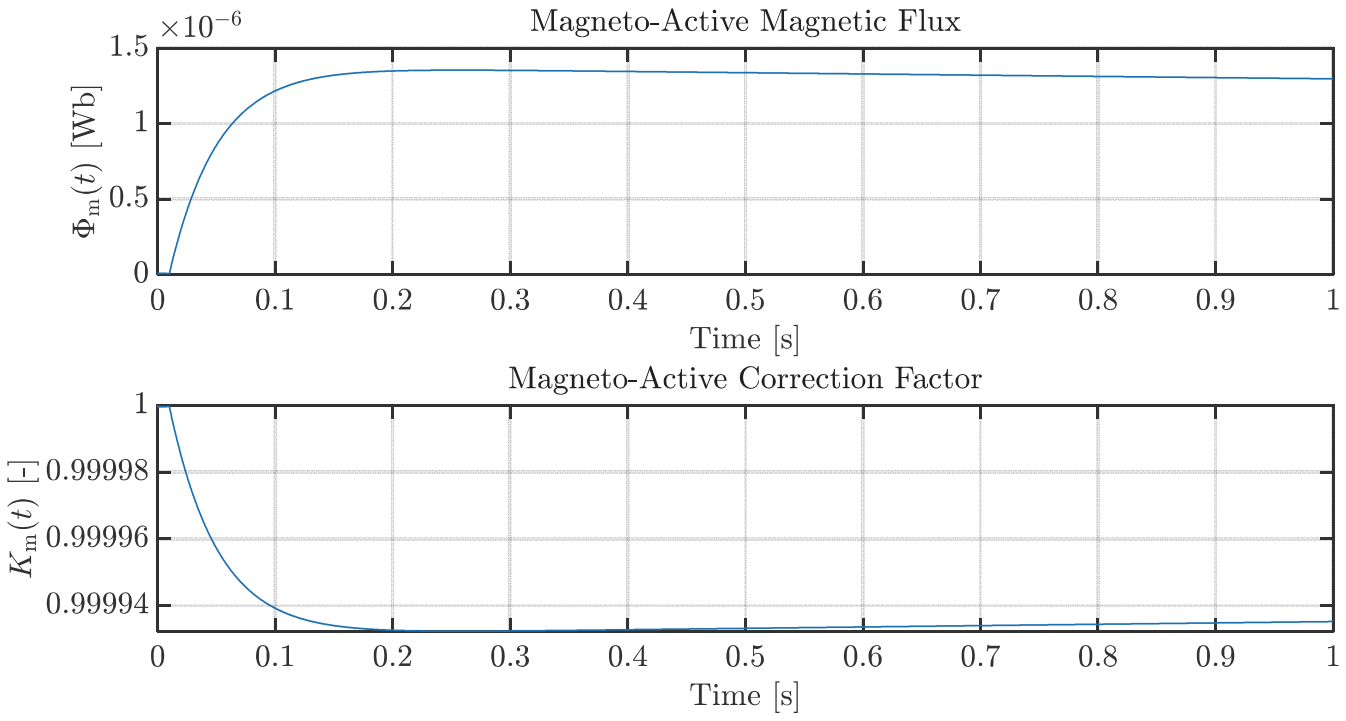

Fig. 3.4. Transient evolution of the magneto-active magnetic flux $\Phi_m(t)$ and magneto-active correction factor $K_m(t)$ associated with the proposed grounding electrode system. The results demonstrate the dynamic electromagnetic coupling behavior introduced by the magneto-active subsystem, including the stabilization of the magnetic flux and the corresponding adaptive modification of the equivalent transient grounding response during fault excitation conditions.

Finally, Fig. 3.5 presents the transient evolution of the ionization correction factor and soil electric field. The electric field surrounding the grounding electrode increases rapidly during the transient interval due to the rise in grounding potential. Nevertheless, the obtained electric field magnitude remains below the critical ionization threshold defined in the proposed model. Consequently, the ionization correction factor remains approximately constant and equal to unity during the entire simulation interval.

This result indicates that the considered transient excitation is insufficient to trigger significant soil ionization phenomena. From a physical perspective, this behavior is entirely coherent, since moderate transient disturbances do not necessarily produce the extremely high electric field intensities required for substantial conductivity enhancement through ionization mechanisms. Therefore, the obtained results validate the consistency of the proposed ionization model while simultaneously confirming that the grounding system remains within its non-ionized operating regime under the considered disturbance conditions.

Overall, the obtained simulation results demonstrate that the proposed MAGE system exhibits stable nonlinear behavior under transient operating conditions. The numerical results confirm the dynamic interaction among electromagnetic propagation, conductive adaptation, thermal diffusion, moisture transport, and magneto-active coupling phenomena. Furthermore, the simulations demonstrate that the grounding system cannot be accurately represented through conventional static grounding resistance formulations alone, since the equivalent grounding impedance evolves dynamically according to the coupled environmental and electromagnetic conditions surrounding the electrode structure.

## V. CONCLUSIONS

This work presented a comprehensive nonlinear dynamic modeling framework for a MAGE system considering the coupled interaction among electromagnetic propagation, grounding impedance dynamics, conductive adaptation, thermal diffusion, moisture transport, and soil ionization phenomena. Unlike conventional grounding analyses based exclusively on static grounding resistance formulations, the proposed approach modeled the grounding structure as a distributed dynamic subsystem whose transient behavior evolves according to environmental conditions, electromagnetic interactions, and conductive adaptation mechanisms occurring within the soil–electrode interface region.

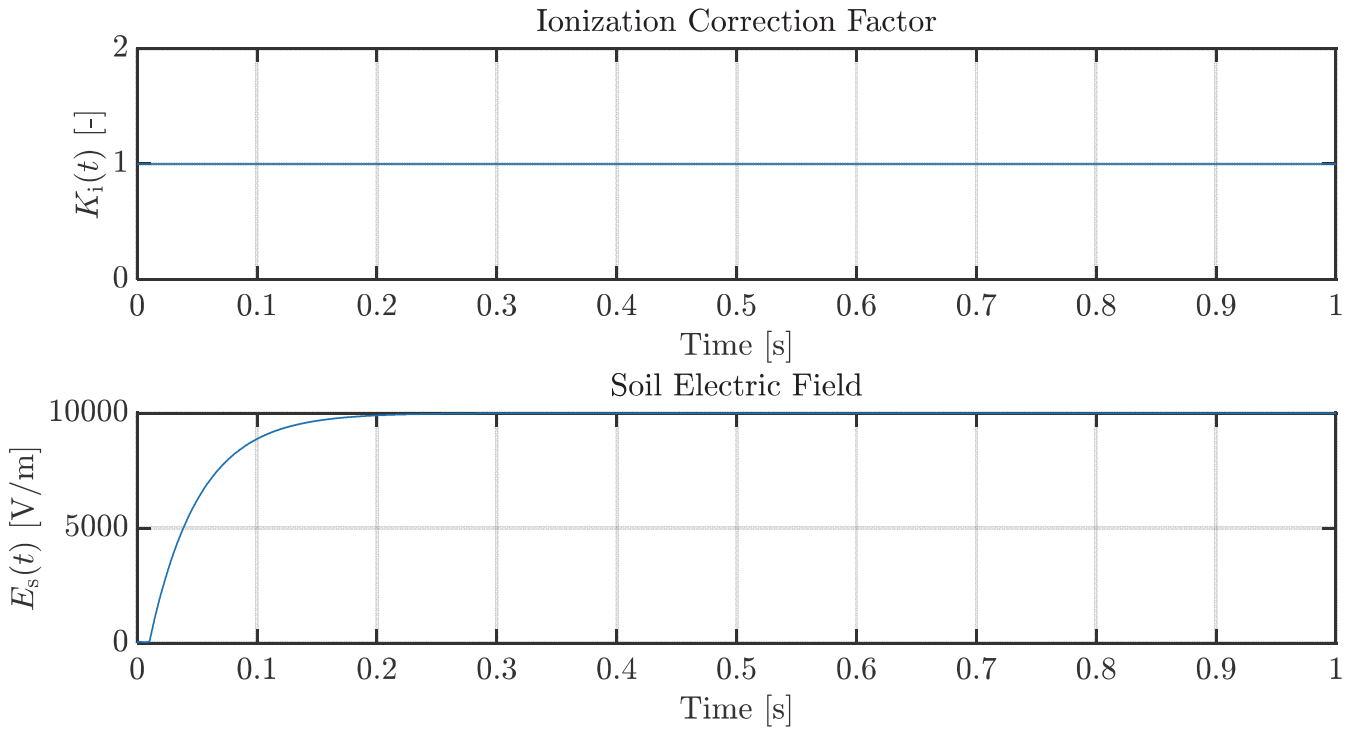

Fig. 3.5. Transient behavior of the ionization correction factor $K_i(t)$ and soil electric field $E_s(t)$ associated with the proposed magneto-active grounding electrode system during electromagnetic disturbance conditions. The results illustrate the evolution of the electric field surrounding the electrode and the corresponding nonlinear ionization response of the soil medium, demonstrating that the operating conditions remain below the critical ionization threshold throughout the simulated transient interval.

The developed formulation incorporated transient electromagnetic equations, nonlinear conductivity dynamics, electrothermal coupling, moisture-dependent soil behavior, magnetic coupling effects, and ionization mechanisms into a unified mathematical framework implemented in the MATLAB/Simulink environment. The resulting model enabled the simultaneous representation of conductive, electromagnetic, thermal, and hydraulic interactions associated with transient current dissipation through the grounding topology.

The obtained numerical results demonstrated that the proposed MAGE system exhibits inherently nonlinear and dynamically coupled behavior under transient operating conditions. In particular, the simulations confirmed that the equivalent grounding impedance cannot be accurately interpreted as a constant parameter, since the grounding resistance dynamically evolves according to conductivity enhancement, thermal diffusion, moisture redistribution, and magneto-active electromagnetic interaction effects. Consequently, the proposed results reinforce the limitations associated with conventional static grounding formulations when applied to transient electromagnetic studies involving advanced grounding technologies.

The transient responses associated with the grounding current and grounding potential rise demonstrated physically coherent electromagnetic behavior characterized by stable transient evolution and asymptotic convergence toward steady-state operating conditions. The absence of numerical oscillations or unstable dynamics additionally confirmed the robustness and internal consistency of the proposed coupled differential-equation framework.

The obtained conductivity and grounding-resistance dynamics further demonstrated the existence of adaptive conductive behavior within the soil–electrode interface region. Specifically, the gradual reduction of the equivalent grounding resistance during the transient process confirmed that electromagnetic coupling, thermal effects, and conductive adaptation mechanisms substantially influence transient grounding performance. This result constitutes one of the principal contributions of the proposed study, since it demonstrates that grounding-system behavior is intrinsically dynamic and environmentally dependent.

The thermal and moisture simulations additionally revealed the strong electrothermal coupling associated with transient current dissipation. The numerical results showed moderate temperature increases caused by Joule heating effects together with slight moisture-content reductions produced by evaporation dynamics. Nevertheless, the conductive backfill subsystem maintained relatively stable moisture-retention characteristics throughout the simulated interval, indicating the potential long-term operational stability of the proposed grounding topology under moderate transient conditions.

The simulations associated with the magneto-active subsystem demonstrated that the integrated electromagnetic module may dynamically influence the transient grounding response through magnetic coupling effects. Although the obtained magnetic correction remained moderate under the considered operating conditions, the results confirmed the capability of the magneto-active assembly to modify the transient electromagnetic behavior of the grounding structure. This observation becomes particularly relevant because conventional grounding models generally neglect the influence of electromagnetic coupling mechanisms integrated directly within grounding electrodes.

Similarly, the ionization analysis demonstrated that the proposed formulation adequately reproduces the nonlinear transition behavior associated with soil electric-field intensification. Under the considered disturbance conditions, the electric field remained below the critical ionization threshold, resulting in negligible ionization enhancement. Nevertheless, the implemented ionization framework establishes a suitable basis for future investigations involving high-energy lightning discharges and extreme transient operating conditions where nonlinear ionization effects may become dominant.

One of the most important scientific contributions of this work corresponds to the establishment of an initial analytical and computational framework for the transient analysis of magneto-active grounding electrodes. To the best of the authors' knowledge, no previous studies reporting nonlinear dynamic models specifically dedicated to MAGE systems were identified during the literature review process. Consequently, the proposed formulation contributes toward addressing an important scientific gap associated with the modeling and transient analysis of next-generation grounding technologies incorporating electromagnetic coupling subsystems.

It is important to emphasize that the present investigation corresponds exclusively to a theoretical proof-of-concept study based on numerical simulations. Therefore, no experimental validation, laboratory measurements, prototype implementation, or field-testing campaigns were performed within the scope of this manuscript. Accordingly, the obtained results should be interpreted as preliminary theoretical evidence supporting the feasibility and dynamic consistency of the proposed modeling approach rather than as definitive experimental validation of the grounding technology itself.

Despite this limitation, the proposed framework establishes a valuable foundation for future research activities associated with advanced grounding technologies. In particular, future investigations may incorporate:

- experimental validation using laboratory-scale grounding prototypes,
- high-energy impulse-current testing,
- finite-element electromagnetic simulations,

- distributed parameter soil models,
- frequency-dependent magnetic permeability formulations,
- nonlinear magnetic saturation effects,
- electrochemical corrosion dynamics,
- seasonal environmental variations,
- and optimization methodologies for grounding-system design.

Additionally, future studies may investigate the application of the proposed framework to substations, renewable-energy installations, high-voltage transmission systems, industrial facilities, and lightning-protection infrastructures operating under severe electromagnetic environments.

Overall, the obtained results demonstrated that magneto-active grounding systems constitute highly coupled nonlinear structures whose transient behavior is governed by the interaction among conductive, electromagnetic, thermal, and environmental phenomena. Consequently, the proposed modeling methodology may provide an important analytical basis for the future development, optimization, and transient analysis of advanced grounding technologies intended for modern electrical power systems operating under increasingly complex electromagnetic conditions.